\documentclass[preprint,12pt]{elsarticle}

\usepackage[utf8]{inputenc}
\usepackage{xcolor}
\usepackage{lineno}
\usepackage{url}

\usepackage[letterpaper,top=2cm,bottom=2cm,left=2cm,right=2cm,marginparwidth=1.75cm]{geometry}

\usepackage{amsmath}
\usepackage{palatino,graphicx,wrapfig}
\usepackage[small,it]{caption}
\usepackage{amsmath}
\usepackage{amssymb,bm}
\usepackage{parskip}
\usepackage{multirow}

\journal{ }

\begin{document}

\title{Chemical Timescale Effects on Detonation Convergence}
\author[label1]{Shivam Barwey\corref{cor1}}
\author[label2]{Michael Ullman}
\author[label2]{Ral Bielawski}
\author[label2]{Venkat Raman}
\cortext[cor1]{Corresponding author. E-mail address: {sbarwey@anl.gov} (S. Barwey).}
\affiliation[label1]{organization={Transportation and Power Systems Division, Argonne National Laboratory},
            city={Lemont},
            postcode={60439},
            state={IL},
            country={USA}}

\affiliation[label2]{organization={Department of Aerospace Engineering, University of Michigan},
            city={Ann Arbor},
            postcode={48109},
            state={MI},
            country={USA}}

\begin{abstract}
Numerical simulations of detonation-containing flows have emerged as crucial tools for designing next-generation power and propulsion devices. As these tools mature, it is important for the combustion community to properly understand and isolate grid resolution effects when simulating detonations. {To this end, the objective of this work is to provide a comprehensive analysis of the numerical convergence of unsteady detonation simulations, with focus on isolating the impacts of chemical timescale modifications on convergence characteristics in the context of operator splitting}. With the aid of an {AMReX-based adaptive mesh refinement flow solver \cite{sharma2024amrex}—which enables resolutions up to $\mathcal{O}(1000)$ cells-per-induction length—}the convergence analysis is conducted using two kinetics configurations: (1) the simplified three-step Arrhenius-based model mechanism {of Short and Quirk \cite{short1997nonlinear}}, where chemical timescales in the detonation are modified by adjusting activation energies in {the initiation and branching reactions}, and (2) {the detailed hydrogen-air mechanism of Shepherd et al.\ \cite{shepherd_h2air_web}}, where the chemical timescales are adjusted by varying the ambient pressure. The convergence of unsteady self-sustained detonations in one-dimensional channels is then analyzed with reference to steady-state theoretical baseline solutions using these mechanisms. The goal of the analysis is to provide a detailed comparison of the effects of grid resolution on both macroscopic (peak pressures and wave speeds) and microscopic (wave structure) quantities of interest, drawing connections between the deviations from steady-state baselines and minimum chemical timescales. {In particular, chemical timescale reductions were found to have minimal impact on the convergence of macroscopic properties. However, analyses of microscopic convergence trends, particularly in the reaction front location, revealed a key insight: maintaining the {induction time} while eliminating prohibitive chemical timescales through mechanism simplifications and combustion modeling can significantly enhance detonation convergence properties.} {Ultimately, this work uncovers resolution-dependent unsteady detonation convergence regimes and highlights the important role played by not only the chemical timescales, but also the ratio between the chemical timescale and induction time on the numerical convergence of the detonation wave structure.} 

Keywords: Detonations; Convergence; Chemical Kinetics; Chemical Timescales; {Adaptive Mesh Refinement}; {Compressible Reacting Flow}
\end{abstract}

\maketitle

\section{Introduction}
\label{sec:introduction}

{Numerical simulation of detonation waves in both canonical \cite{poludnenko2019unified,wang_2023,meagher2023isolating} and complex configurations \cite{venkat_rde_arfm,supraj_liquid} is becoming increasingly important due to emerging interest in detonation-based propulsion devices \cite{gutmark2019_pecs,wolanski2013_pci}. Such simulations are also crucial in other related applications, including condensed-phase explosives \cite{short_arfm_2018,voelkel2022effect} and hazard prediction \cite{yanez2015analysis,ng2007hazard}. 

Regarding canonical simulations, significant ongoing efforts have been put toward performing detonation simulations with the aim of extracting fundamental physical insights into detonation-flow-chemistry interactions at levels beyond what is achievable with experimental diagnostics. These efforts span from one-dimensional to three-dimensional configurations. For example, one-dimensional numerical studies revolve around investigating and modeling nonlinearities (e.g., chaotic wave propagation behavior and chemistry effects) in detonations using both simplified and detailed chemical mechanisms in highly-resolved steady \cite{powers_paolucci} and unsteady \cite{mazaheri2007effect,ng2005numerical,yungster_ctm_2004,short1997nonlinear,han2019pulsation} settings. 

Building on one-dimensional studies, numerical investigations in higher spatial dimensions have explored the fundamental physical properties of cellular detonation wave structures and their dynamic behavior across a wide range of fuel types and configurations. Earlier studies analyzed the unsteady behavior of detonations under operating conditions relevant to rotating \cite{schwer_kailasanath, mazaheri2012diffusion} and pulse \cite{ebrahimi_pulse} detonation engines. Since then, the rapid increase in computational capability in recent years has enabled detailed investigations of complex multi-physics interactions in canonical detonation-containing flows. Recent examples include, but are not limited to, (1) investigating the effect of liquid fuel injection on the detonation wave structure and particle distributions \cite{musick2023numerical, meng2021distributions}, (2) capturing the effects of turbulence interactions on the detonation wave structure and propagation behavior (e.g., via imposition of fuel stratification \cite{michael_stratification, ryu2024numerical} or turbulence inflow generation \cite{suzuki2024dns}), (3) detailed investigations into the highly transient deflagration-to-detonation phenomenon \cite{machida2014ddt,ramachandran2023ddt}, and (4) resolved simulations of oblique detonations \cite{han2019three, abisleiman2024high}. 

The above canonical simulations are used to inform and supplement full-geometry simulations of detonation-based power and propulsion concepts, including pulse \cite{debnath2023computational,debnath2023numerical}, rotating \cite{takuma_proci,pinaki_les_rde} and linear \cite{ullman2023_drone} detonation engines. In the space of rotating detonation engines specifically, state-of-the-art computational tools can now be used to conduct resolved, full-scale multi-phase detonation simulations of realistic geometries and fuels \cite{supraj_liquid,menon_2024}. Ultimately, in all practical detonation-based applications, detonation dynamics are complicated by counter-propagating waves \cite{ullman2023_drone}, heterogeneous mixtures \cite{supraj_linearized, supraj_rde_proci}, and surface stabilization effects \cite{rosato2021stabilized}, leading to complex detonation structures influenced by fine-scale flow-chemistry interactions that are difficult to probe experimentally due to the extreme conditions and short timescales in these systems \cite{venkat_rde_arfm}.  
}

{Due to the inherent complexity and importance in achieving physically-accurate detonation simulations, significant effort has been devoted to developing robust, predictive, and scalable numerical methods for solving the governing compressible reacting Navier-Stokes equations}. Recent approaches have utilized numerics rooted in finite volume methods for full-geometry \cite{detfoam_paper} and canonical adaptive mesh refinement (AMR)-based simulations \cite{hu2017numerical}. Additionally, higher-order extensions such as the piecewise parabolic method coupled with conservative shock tracking strategies \cite{bourlioux_ppm_1991}, as well as discontinuous Galerkin methods \cite{dg_detonation}, have been effectively used to study fundamental detonation dynamics. Alternative strategies like the ghost fluid method can also be used \cite{fedkiw1999ghost}, which resolves shock-chemistry interactions through level set equations. 

The manner in which chemical reactions are accounted for during the flow evolution is crucial to all of these methods. In the level set approaches, chemistry coupling is accounted for through models for front propagation speeds \cite{fedkiw1999ghost}. An alternative approach is operator splitting \cite{detfoam_paper}, which decouples the effect of chemical reactions from advection- and diffusion-based transport, allowing one to invoke dedicated stiff solvers to treat chemistry contributions. Strategies that directly couple chemistry and transport have been recently explored to mitigate splitting-related issues \cite{blanquart_jcp_2015}, although the added computational overhead typically prevents such methods from being used in full-scale detonation simulations requiring detailed chemistry descriptions.

Regardless of the approach used, proper understanding of grid convergence properties in PDE-based simulation of detonating flows is of utmost importance. Analysis of detonation convergence properties to this end has primary focused on identifying onsets of instability, with observations showing how acoustic waves generated in the induction zone at high resolutions—which would otherwise have been implicitly filtered at low resolutions—lead to detonation quenching behavior \cite{qian2020convergence,mazaheri_thesis}. Additionally, convergence analysis performed in Ref.~\cite{yungster2005structure} for ethylene detonations reports small variation in mean wave speeds, as well as the emergence of high-frequency and low-amplitude propagation modes in highly resolved cases. Similar behavior was observed in the context of n-Heptane detonations \cite{zhao2021pulsating}, where an initial convergence analysis step was carried out to isolate resolution effects on pulsating frequencies.

These previous works have investigated resolution effects in the context of physical analysis of detonation instability onsets and behavior. While important, this naturally results in less emphasis on direct and rigorous analysis of detonation convergence itself; even in stabilized detonation configurations, a dedicated convergence analysis can yield valuable physical and numerics-oriented insights. Additionally, although connections between limiting spatial resolutions and chemical kinetics in detonations have been made in the context of steady-state ODE models \cite{powers_paolucci}, the impact of chemical timescales on the detonation structure and their influence on detonation convergence properties have not been explored for PDE-based solutions.

As such, the goal of this work is to investigate the impact of chemical timescale modifications on the numerical convergence of unsteady detonations using simulations based on operator splitting. The objective is both to conduct a comprehensive detonation convergence analysis, and to isolate the chemical timescale as a tunable parameter from two different angles: (1) through direct modification of reaction mechanism parameters (i.e., activation energies), and (2) through modification of the simulation operating conditions (ambient pressures). The former leverages a simplified model three-step mechanism \cite{short1997nonlinear} and the latter leverages a detailed hydrogen mechanism \cite{shepherd_h2air_web}. {To efficiently conduct numerical convergence studies, this work leverages a recently-developed AMReX-based flow solver \cite{sharma2024amrex}, which discretizes the compressible reacting Navier-Stokes equations using a block-structured adaptive mesh refinement strategy.} With this solver, simulation convergence trends in terms of quantities of interest (QoIs) relevant to detonation engine design, such as wave speeds, peak pressures, and detonation wave structures, are analyzed using highly resolved simulations of unsteady detonations. {The specific contributions of this work are provided below.
\begin{itemize}
    \item Detonation convergence analyses are performed using both simplified and detailed chemistry mechanisms in the AMR-based flow solver. The effects of chemical timescales on convergence are accessed through (a) activation energy modifications in the simplified mechanism, and (b) ambient pressure modifications in the detailed mechanism.
    
    \item Where applicable, comparisons are made to steady-state Zeldovich-von Neumann-D\"{o}ring (ZND) baseline solutions, such that the effects of chemical timescale modifications on convergence trends can be monitored through deviations of QoIs from ZND solutions in highly-resolved regimes.
    
    \item Convergence analysis is performed for both macroscopic QoIs and the fine-scale detonation wave structure. The wave-structure analysis reveals three key convergence regimes: an under-resolved regime, a strongly-coupled regime characterized by shortened induction zones, and a ZND-conforming regime. 

\end{itemize}}

{The paper proceeds as follows. In Sec.~\ref{sec:numerics}, the numerical approach is outlined, and details are provided for the chemical mechanisms and the processes by which the chemical timescales are modified. Results are provided in Sec.~\ref{sec:results}, with Sec.~\ref{sec:znd} outlining the steady-state solution baselines, Sec.~\ref{sec:macro} detailing the convergence trends of macroscopic detonation QoIs (e.g., peak pressures and wave speeds), and Sec.~\ref{sec:wave_structure} providing convergence trends in terms of the fine-scale detonation wave structure. Concluding remarks are provided in Sec.~\ref{sec:conclusion}.}

\section{Numerical Approach}
\label{sec:numerics}
{The governing equations are the compressible reacting Navier-Stokes equations:
\begin{equation}
    \label{eq:mass}
    \frac{\partial \rho}{\partial t} + \frac{\partial \rho u_i}{\partial x_i} = 0,
\end{equation}
\begin{equation}
    \label{eq:momentum}
    \frac{\partial \rho u_j}{\partial t} + \frac{\partial \rho u_i u_j}{\partial x_i} = -\frac{\partial p}{\partial x_j} + \frac{\partial \tau_{ij}}{\partial x_i},
\end{equation}
\begin{equation}
    \label{eq:energy}
    \frac{\partial \rho E}{\partial t} + \frac{\partial \rho u_i H}{\partial x_i} = \frac{\partial}{\partial x_i} \left( \alpha \frac{\partial T}{\partial x_i} \right) + \frac{\partial \tau_{ij} u_i}{\partial x_j} + \sum_k \left( h_k \frac{\partial}{\partial x_i} \left( \rho D \frac{\partial Y_k}{\partial x_i}\right) \right),
\end{equation}
\begin{equation}
    \label{eq:species}
    \frac{\partial \rho Y_k}{\partial t} + \frac{\partial \rho u_i Y_k}{\partial x_i} = \frac{\partial}{\partial x_i} \left( \rho D \frac{\partial Y_k}{\partial x_i} \right) + \dot{\Omega}_k,
\end{equation}
where
\begin{equation}
    \label{eq:tau_ij}
    \tau_{ij} = -\frac{2}{3}\mu \frac{\partial u_k}{\partial x_k}\delta_{ij} + \mu \left( \frac{\partial u_j}{\partial x_i} + \frac{\partial u_i}{\partial x_j} \right).
\end{equation}
Equations \ref{eq:mass}-\ref{eq:species} respectively correspond to mass, momentum, energy, and species conservation. Here, $\rho$ is the fluid density, $t$ is time, $x_i$ is the spatial position, $u_i$ is the fluid velocity, $p$ is the fluid pressure, $\tau_{ij}$ is the viscous stress tensor, $\mu$ is the dynamic viscosity, $E$ is the total chemical energy, $H$ is the total mixture enthalpy, $h_k$ is the species enthalpy (including sensible and chemical components), $\alpha$ is the thermal conductivity, $Y_k$ is the mass fraction of the $k$-th species, $D$ is the local diffusivity (assumed equal for all species), and $\dot{\Omega}_k$ is the chemical source term for species $k$. In this work, viscosity and diffusion effects are neglected in the governing equations for both the simplified and detailed chemical mechanisms. An ideal gas equation of state is nominally used to close the system. In the simplified mechanism detailed in Sec.~\ref{sec:simple_chem}, the gas is further assumed to be calorically perfect.}

{In this work, the above equations are solved in one spatial dimension} and are discretized in time using a globally second-order Strang splitting approach \cite{strang}, which decouples the effects of chemical reactions from those of fluxes. The strategy used here is the routinely used ``Reaction$\rightarrow$Advection/Diffusion$\rightarrow$Reaction" order of operator invocations. As such, there are two chemical time integration steps in a single global time step advance. 

{The equations are solved using an in-house solver based on the AMReX framework \cite{sharma2024amrex}.} The solver is a block-structured adaptive mesh refinement (AMR) extension of the extensively verified UMReactingFlow \cite{detfoam_paper}. {On each mesh refinement level}, chemical time integration is handled using an adaptive explicit method, and the advection-diffusion advance is treated using a stability-preserving second-order Runge-Kutta method combined with a slope-limited approximate Riemann solver. This work specifically leverages a van Leer limiter with the Harten–Lax–van Leer contact (HLLC) approximate Riemann solver \cite{hllc}. {Grid refinement is handled by AMReX \cite{AMReX_JOSS}, which provides a backend for the grid communication steps used in the block-structured AMR formulation.} The reader is directed to Ref.~\cite{sharma2024amrex} for additional detail on the numerics of each component and the solution procedure. 

\begin{figure}
    \centering
    \includegraphics[width=0.75\textwidth]{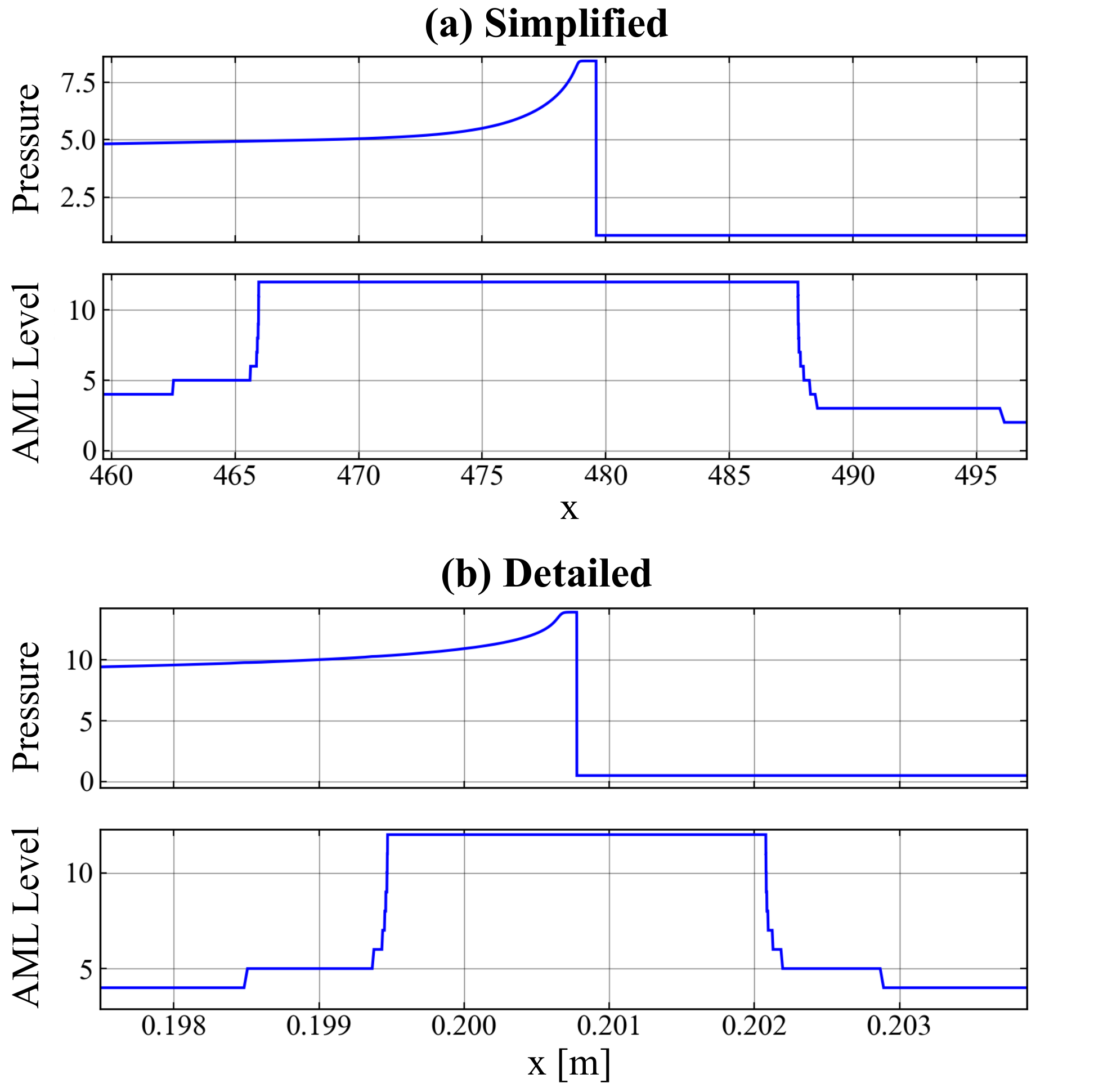}
    \caption{{\textbf{(a)} Snapshots of pressure (top) and AMR level (bottom) profiles near detonation wave for a 12-level simulation in the Simplified Case 2 configuration (see Table~\ref{table:cases}) during unsteady detonation propagation. \textbf{(b)} Same as (a), but for the Detailed Case 2 configuration.}}
    \label{fig:pressure_level_profiles}
\end{figure}

The AMR tagging criterion is based on pressure and density gradients with additional buffer cell padding, and was determined to properly isolate the detonation wave structures for subsequent mesh refinement. Grid refinement factors were set to 2, and all convergence resolutions reported in Sec.~\ref{sec:results} are extracted from the finest (highest) AMR level, with the highest level count being 14. Separate verification studies were conducted to ensure that AMR-induced interpolation errors did not pollute detonation profiles for the range of levels considered. Examples of unsteady detonation profiles showcasing AMR level coverage in a 12-level case for both simplified (Sec.~\ref{sec:simple_chem}) and detailed (Sec.~\ref{sec:detailed_chem}) chemistry configurations is shown in Fig.~\ref{fig:pressure_level_profiles}. The figure shows that the refinement criteria used in the AMR simulations successfully resolve the detonation profile at the desired level specification.

The solver is used to evaluate chemical timescale effects on the convergence of one-dimensional unsteady detonations, leveraging both simplified and detailed chemical kinetics configurations. Descriptions of each configuration are provided below. In all cases, the computational domain consists of a wall at the left boundary and a zero-gradient outlet at the right. Detonations are initiated in simplified chemistry cases using high-energy driver gases in a small region at the left wall, while ZND solutions are used to initialize detailed chemistry cases. Overdrive effects from the wall during initiation were found to not affect solutions. Convergence trends reported in this study come from post-initiation regimes in which detonations are self-sustained.

\subsection{Simplified Chemistry}
\label{sec:simple_chem}
The simplified chemistry description is the three-step Arrhenius-based model mechanism of Ref.~\cite{short1997nonlinear}. The model mechanism is given by

\begin{equation} 
\label{eq:model_mechanism}
    \begin{split}
    [F] & \overset{k_{I}}\longrightarrow [R], 
    \quad k_I = \exp\left(E_I \left( \frac{1}{T_I} - \frac{1}{T}\right)\right), \\
    [F] + [R] & \overset{k_{B}}\longrightarrow 2[R], 
    \text{  } k_B = \exp\left(E_B \left( \frac{1}{T_B} - \frac{1}{T}\right)\right), \\ 
    [R] & \overset{k_C}\longrightarrow [P], \quad k_C = 1.
    \end{split}
\end{equation}

The mechanism consists of temperature-dependent, radical-producing chain initiation and chain branching forward reactions (indicated by the reaction rates $k_I$ and $k_B$), followed by a temperature-independent termination reaction $k_C$ at unity rate. The quantities $[F]$, $[R]$, and $[P]$ denote fictitious fuel, radical, and product species concentrations, respectively.

There are four key parameters: the initiation and branching activation energies ($E_I$ and $E_B$), and the corresponding crossover temperatures ($T_I$ and $T_B$). This work leverages the activation energies to control the sensitivity of the detonating flow field to chemical reactions, providing a direct pathway to parameterize chemical timescales within the detonation wave structure, as showcased in Sec.~\ref{sec:znd}. In particular, three different activation energy settings (Cases 1-3 in Table~\ref{table:cases}) were prescribed to explore a wide range of chemical timescale influence in the simplified configurations. Besides the activation energies, all other parameters match those used for the lowest $T_B$ case in Ref.~\cite{ng2003direct}. To admit a theoretically valid detonation solution with this mechanism, the gas is assumed to be calorically perfect with constant specific heat ratio $\gamma$, leading to a simplified definition of chemical energy. This differs from the detailed case, where variable $\gamma$ is used. All length and time values in the simplified configurations are reported in non-dimensional quantities, where non-dimensionalization is carried out such that the termination rate $k_C$ is unity. The reader is directed to Ref.~\cite{ng2003direct} for additional details. 

The length of the one-dimensional domain for the simplified cases was 800 units with a uniform base grid consisting of 800 cells, yielding a base grid resolution of 1 distance unit. The minimum grid size in the finest simulation is roughly $1.22\times10^{-4}$ distance units. To initialize unsteady simulations, a driver gas at elevated pressure and temperature with uniform profile in the range $x = [0,10]$ was imposed at the left-most wall as the initial condition. The pressure and temperature ratios of the driver gas were set to $P_D/P_0 = 20$ and $T_D/T_0=7$ respectively, with the driver gas composition given by 100\% product mass fraction.

\subsection{Detailed Chemistry}
\label{sec:detailed_chem}
Detailed chemical kinetics are modeled using the 14 species, 42 reaction H$_2$-air mechanism of Shepherd et al.\ \cite{shepherd_h2air_web}, which is based on the mechanism of M{\'e}vel et al.\ \cite{shepherd_h2air_3} \footnote{{The mechanism is openly available at the following link: \url{https://shepherd.caltech.edu/EDL/PublicResources/sdt/SDToolbox/cti/Mevel2017.cti}}}. The one-dimensional domain was 0.8 m long with a uniform base grid of 1600 cells, yielding a base resolution of 500 $\mu$m. The minimum grid size in the finest simulation is 30.5 nm. All cases used the reactant molar composition 2H$_2$:O$_2$:7Ar at an ambient temperature of 300 K. Significant argon dilution was added to ensure stability of the detonation waves \cite{radulescu2002_argon}. In order to explore different chemical timescale regimes, ambient pressures of 0.2, 0.5, and 1 atm were investigated (as described in Table~\ref{table:cases}). For each case, the ZND solution was computed using the Shock and Detonation Toolbox \cite{sdtoolbox}. To initialize unsteady simulations, this solution was then mapped onto the computational grid at the initial time, with the wave front positioned 0.2 m from the left boundary (i.e., the driver gas was the ZND wave in the detailed cases, as these were found to more quickly lead to self-sustained detonations). 

\begin{table}[h]
\centering
\renewcommand{\arraystretch}{1.} 
\setlength{\tabcolsep}{15pt} 
\begin{tabular}{|c|p{10cm}|} 
\hline
\textbf{Symbol} & \textbf{Description} \\
\hline
$E_I, E_B$ & Initiation and branching activation energies used in the simplified mechanism. \\
\hline
$T_I, T_B$ & Initiation and branching crossover temperatures using in the simplified mechanism. \\
\hline
$c_0$ & Ambient speed-of-sound. \\
\hline
$P_0$ & Ambient pressure. \\
\hline
$\tau_{\text{chem}}$ & Minimum chemical timescale observed in the detonation wave, obtained from the largest eigenvalue of the chemical source term Jacobian. \\
\hline
$\tau_{\text{ind}}$ & Induction time. \\
\hline
$L_{\text{ind}}$ & Induction zone length. \\
\hline
$L_{\text{react}}$ & Reaction zone length. \\
\hline
$W$ & Detonation wave speed observed from numerical simulations. \\
\hline
$W_{\text{CJ}}$ & Wave speed of Chapman-Jouguet detonation. \\
\hline
$P_{\text{max}}$ & Maximum pressure observed on the detonation wave. \\
\hline
$P_{\text{VN}}$ & Von-Neumann pressure observed in the theoretical ZND solution. \\
\hline
$x_s$ & Spatial position relative to leading shockwave. \\
\hline
\end{tabular}
\caption{{Reference nomenclature table for relevant symbols and variables used throughout this paper.}}
\label{tab:nomenclature}
\end{table}

\section{Results \label{sec:results}}

\begin{table*}[ht]
\centering
\footnotesize 
\begin{tabular}{|c|c|c|c|c|c|c|} 
\hline

\multirow{2}{*}
    {
        \begin{tabular}{c}
           Case\\Description
        \end{tabular}
    } & \multicolumn{3}{c|}{\textbf{Simplified Kinetics (Sec.~\ref{sec:simple_chem})}} & \multicolumn{3}{c|}{\textbf{Detailed Kinetics (Sec.~\ref{sec:detailed_chem})}}\\ 
\cline{2-7} 
 & \textbf{Case 1} & \textbf{Case 2} & \textbf{Case 3} & \textbf{Case 1} & \textbf{Case 2} & \textbf{Case 3} \\ 
\hline
    $(E_I/c_0^2, E_B/c_0^2)$ & (37.5, 10) & (82, 25) & (100, 35) & -- & -- & -- \\ 
    $P_0$ (atm) & -- & -- & -- & 0.2 & 0.5 & 1 \\
    $\tau_{\text{chem}}$ & 3.396 $\times 10^{-2}$ & 2.972 $\times 10^{-4}$ & 9.472  $\times 10^{-6}$ & 3.362 $\times 10^{-10}$~s & 1.271 $\times 10^{-10}$~s & 6.100 $\times 10^{-11}$~s  \\
    {$\tau_{\text{chem}}/\tau_{\text{ind}}$ } & {2.333 $\times 10^{-2}$} & {3.645 $\times 10^{-4}$} & {1.033  $\times 10^{-5}$ } & {3.680 $\times 10^{-4}$ } & {3.985 $\times 10^{-4}$ }&  {4.242 $\times 10^{-4}$  } \\
\hline
\end{tabular}
\caption{Detonation case configurations. First row: activation energies in simplified mechanism normalized by squared speed-of-sound. Second row: ambient pressures. Third row: minimum chemical timescale observed in steady-state solution. Fourth row: Ratio of minimum chemical timescale and {induction time} in steady-state solution. {Induction time defined as $\tau_{\text{ind}}=\int_0^{L_\text{ind}} \frac{1}{U(x_s)} \text{d}x_s$, where $L_{\text{ind}}$ is induction length, $x_s$ distance behind the leading shock, and $U(x_s)$ is the local fluid velocity in the shock reference frame}. ZND profiles shown in Fig.~\ref{fig:znd_comparison}.}
\label{table:cases} 
\end{table*}

Convergence of unsteady detonation simulations is analyzed with reference to steady-state ZND solutions using the mechanisms described above. The goal of the analysis is two-fold. First, present a set of case configuration options for both simplified and detailed mechanisms that isolate the effect of limiting chemical timescales in the steady (reference) detonation wave structure (Sec.~\ref{sec:znd}). Then, conduct unsteady AMR-based simulations on the different configuration options, providing a detailed comparison of the effect of grid resolution on both macroscopic (i.e., peak pressures and detonation wave speeds, Sec.~\ref{sec:macro}) and microscopic (detonation wave structure, Sec.~\ref{sec:wave_structure}) quantities of interest. Connections are then drawn between the deviations of self-sustained unsteady detonations from the steady-state baselines and the minimum chemical timescales observed in the detonation.  

\subsection{Case Setup and Steady-State Baselines}
\label{sec:znd}

Three cases (1-3), described in Table~\ref{table:cases}, are used in each of the simplified and detailed configurations, with increase in case number corresponding to decrease in minimum chemical timescale in the detonation. These cases admit steady-state solutions that will serve here as baselines for the unsteady simulations. It should be noted that certain key parameters, such as the branching crossover temperature in the simplified cases and the mole fraction of Argon diluent in ambient gas for the detailed cases, were chosen to ensure \textit{stable} detonation wave propagation in unsteady simulations such that comparisons to the reference ZND wave structures are well-grounded.

\begin{figure}[t]
    \centering
    \includegraphics[width=0.7\columnwidth]{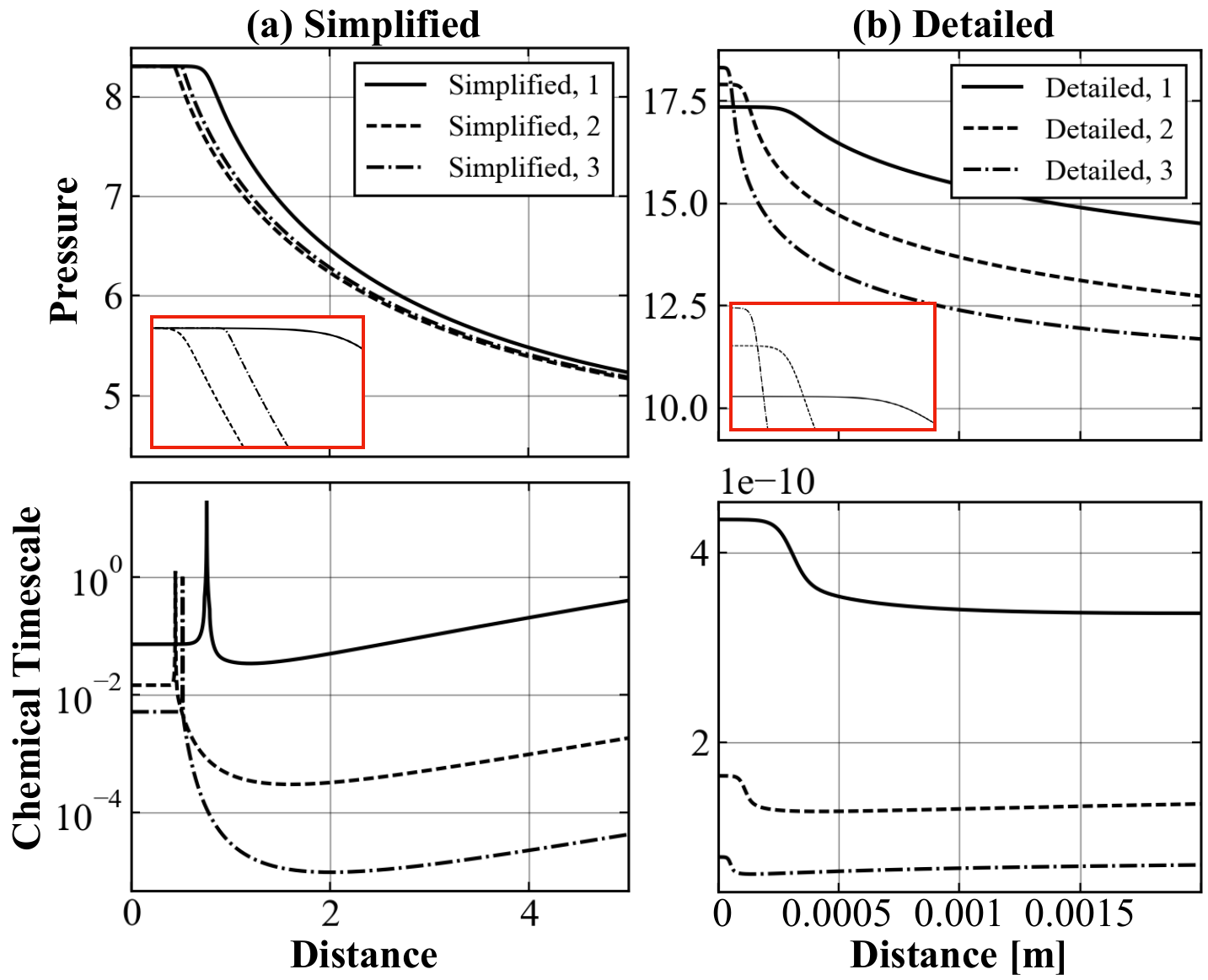}
    \caption{\textbf{(a)} Steady-state (ZND) profiles for simplified kinetics cases outlined in Table~\ref{table:cases}. Top row shows pressure normalized by $\gamma_0 P_0$, where $\gamma_0$ is ambient specific heat ratio and $P_0$ is ambient pressure (red insets show zoom-ins near induction zone). Bottom row shows minimum chemical timescale, computed as $\text{min}(1/|\lambda_i|)$, where $\lambda_i$ is an eigenvalue of chemical source term Jacobian. \textbf{(b)} Same as (a), but for detailed kinetics cases.}
    \label{fig:znd_comparison}
\end{figure}

Steady-state ZND profiles for both simplified and detailed configurations are shown in Fig.~\ref{fig:znd_comparison}. The figure displays reference wave structures in the form of normalized pressures and limiting chemical timescales corresponding to the case descriptions provided in Table~\ref{table:cases}. 

Reference solutions for the simplified three-step kinetics description are shown in Fig.~\ref{fig:znd_comparison}(a), where variation in Cases 1 to 3 come from successive increases to both initiation and branching activation energies. Increasing these activation energies results in significant reduction of the minimum chemical timescale encountered within the detonation profile \textit{without significantly altering the detonation wave structure}, thereby providing an effective knob for the chemical stiffness encountered within the detonation reaction zone. This quality is a direct consequence of the simplicity of the three-step mechanism. Since (a) initiation and branching reaction steps influence radical species production only, and (b) the model for chemical heat release is a function of the product mass fraction only \cite{ng2003direct}, alterations to the activation energies provide direct and interpretable access to the sensitivities of chemistry (the timescales) on the flow. When moving between case 1 to 2 (and also 2 to 3), the minimum timescales drop by roughly two orders of magnitude.

Due to the complexity in high-fidelity kinetics representations, isolating activation energies in the same manner as with the simplified model is non-trivial. As a workaround, instead of activation energy modifications, the core parameter for moving between configurations 1 to 3 in the detailed mechanisms is the ambient pressure $P_0$ (see Table~\ref{table:cases}). The corresponding steady-state profiles are shown in Fig.~\ref{fig:znd_comparison}(b).

The ambient pressure modifications result in much more variation in both the von-Neumann (VN) pressures and the steady wave structure as compared to the simplified mechanism counterparts. However, increases in the ambient pressure achieve an analogous reduction to chemical timescales in the wave structure as observed in the simplified case, although the reduction factor in the limiting chemical timescale is much less pronounced.

\begin{figure}
  \centering
  \includegraphics[width=0.6\textwidth]{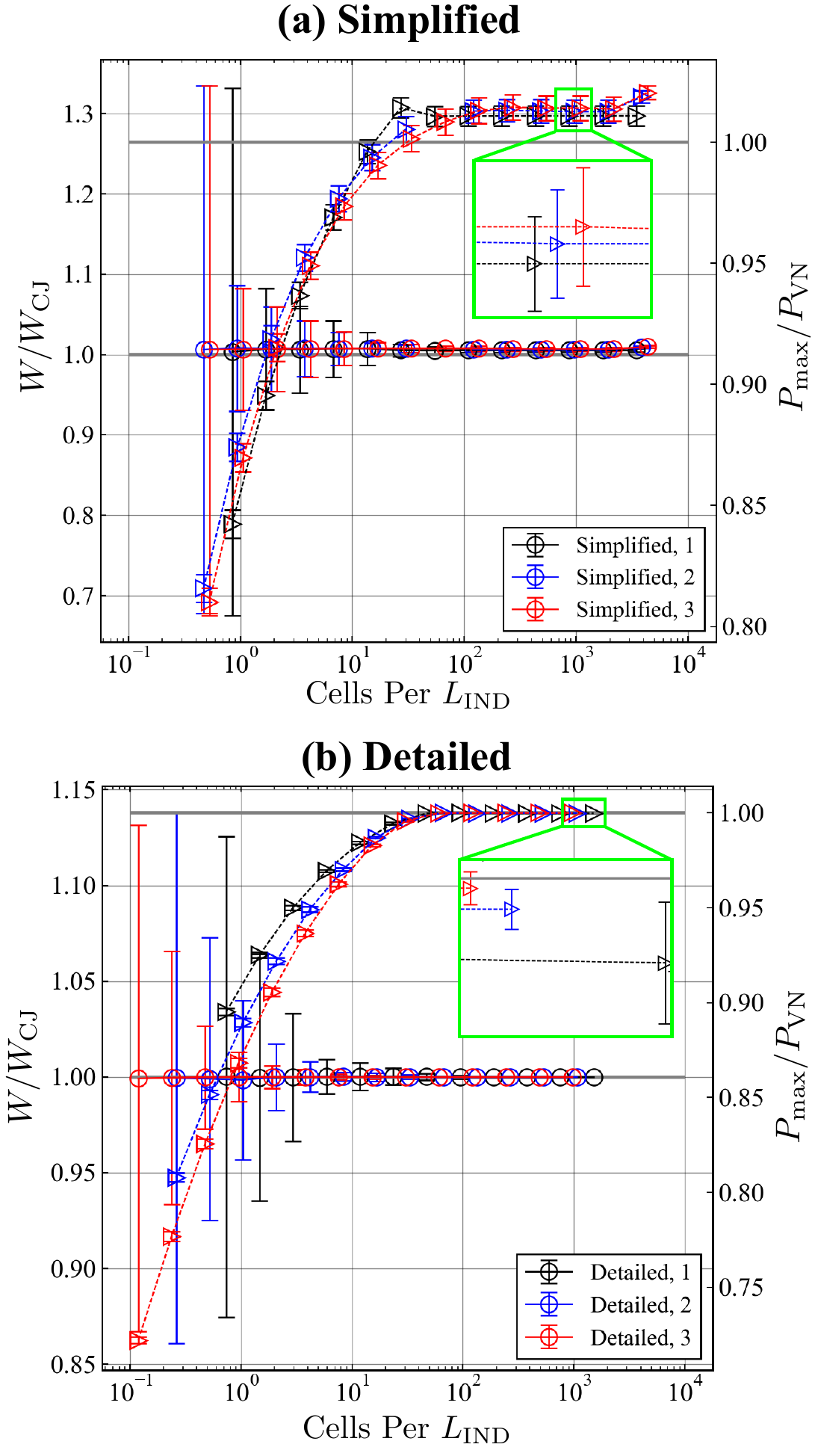}
  \caption{{\textbf{(a)} Detonation wave speeds (circles, normalized by CJ speed) and peak pressures (triangles, normalized by VN pressure) versus grid resolution expressed as cells-per-$L_{\text{IND}}$ in the simplified kinetics configuration for case 1 (black), 2 (blue), and 3 (red). Markers denote mean values extracted during unsteady self-sustained wave propagation, and error bars denote $\pm$ one standard deviation. \textbf{(b)} Same as (a), but for detailed kinetics configuration. Reader is referred to Table~\ref{table:cases} for case details.}}
  \label{fig:w_p_convergence}
\end{figure}

Ultimately, when moving from cases 1 to 3, chemical timescales are being reduced in both simplified and detailed kinetics configurations. Since stable unsteady simulations are expected to relax to the steady profiles in the shock-reference frame, the adjustment of activation energy within the model mechanism and ambient pressure within the detailed mechanism provides a qualitatively comparable way to decrease the chemical timescales. 

\subsection{Convergence of Macroscopic QoIs}
\label{sec:macro}

Figure~\ref{fig:w_p_convergence} shows numerical convergence trends for peak pressures and detonation wave speeds for the simplified and detailed configurations outlined in Table~\ref{table:cases}. For monitoring relative convergence trends, the grid resolution is described in terms of cells-per-$L_{\text{IND}}$, where $L_{\text{IND}}$ represents the induction length defined here as the distance from the initial shock to the point of maximum chemical heat release rate encountered in the steady-state solutions.

In the simplified kinetics description in Fig.~\ref{fig:w_p_convergence}(a), the peak pressures for all three cases achieve maximum values at roughly 20-to-50 cells-per-$L_{\text{IND}}$, and enter a pressure-converged regime at nearly 100 cells-per-$L_{\text{IND}}$. The under-resolved regime below this point is characterized by "weakened" detonations that are still self-sustained, but exhibit markedly decreased shock pressures relative to the baseline ZND solution. On the other hand, in the converged regime beyond 100 cells-per-$L_{\text{IND}}$, the peak pressures stabilize at values slightly above the VN point, indicating a small degree of wave overdrive that is also reflected in the converged wave speeds. 

The peak pressure convergence trends in the detailed kinetics configurations in Fig.~\ref{fig:w_p_convergence}(b) are qualitatively very similar to their simplified kinetics counterparts. Specifically, convergence to the peak value is encountered at nearly the same point (20-50 cells), and the behavior before this point in the highly under-resolved regime is again characterized by decreased pressure values relative to the ZND baselines. 

Convergence in wave speeds is also qualitatively similar across simplified and detailed configurations, with trends markedly different from peak pressure convergence histories. The wave speeds converge almost exactly to the CJ detonation in the detailed cases with no overdrive (Fig.~\ref{fig:w_p_convergence}(b)), and converge to overdriven detonations in the simplified cases (Fig.~\ref{fig:w_p_convergence}(a)). Most notably, wave speed convergence is characterized by negligible changes in the mean values, but drastic reductions in standard deviations with increased grid resolution. The reduction in wave speed variation coincides with convergence of peak pressures to the VN point, and signifies a stabilization of the self-sustained detonations. The large variations in wave speed in the under-resolved simulations are due to the diffused nature of the detonation front over many grid cells. Although this variation is not present in peak pressure values, this phenomenon can be traced to analogous fluctuations in the peak radical mass fractions, the convergence of which showcases an interesting blend of both wave speed and pressure curves (not shown here).

\begin{figure}
  \centering
  \includegraphics[width=\textwidth]{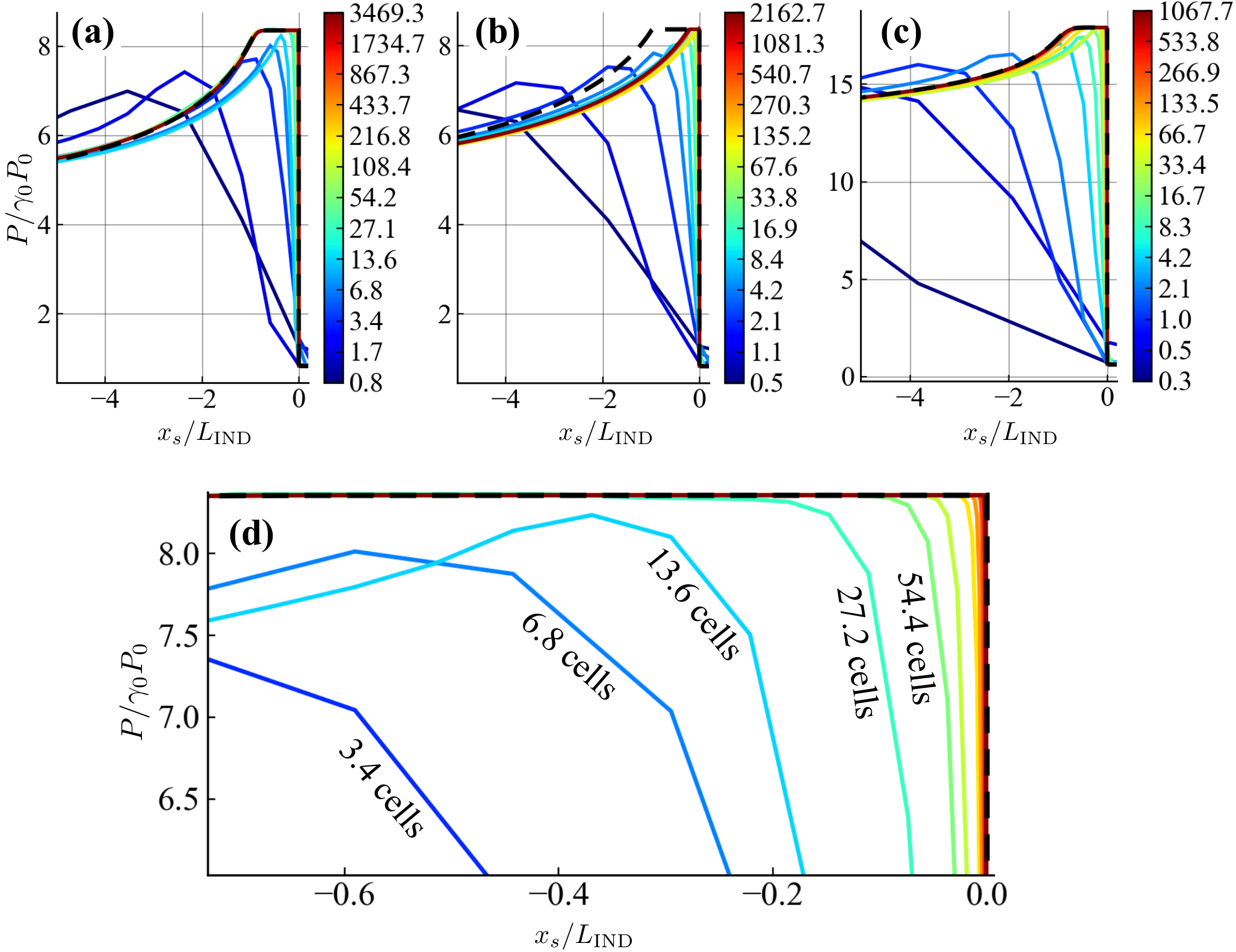}
  \caption{{\textbf{(a)} Instantaneous pressure (normalized by ambient value) versus relative position from shock front (normalized by the ZND induction length, $L_{\text{IND}}$) for simplified Case 1 configuration (refer to Table~\ref{table:cases} for case information). Curves correspond to simulations at increasing cells-per-$L_{\text{IND}}$ as per the color bar. Black dashed line corresponds to ZND solution (in plots, ZND peak pressures are scaled to match converged cases to eliminate overdrive effects for visualization purposes). \textbf{(b)} Same as (a), but for Case 3 in the simplified configuration. \textbf{(c)} Same as (a) and (b), but for Case 2 in the detailed kinetics configuration. \textbf{(d)} Zoom-in of pressure profiles from (a) near shock front, showing transition from strongly-coupled to ZND-conforming regime.}}
  \label{fig:pressure_profiles}
\end{figure}

Recall that in both the simplified and detailed descriptions outlined in Sec.~\ref{sec:znd}, the variations between cases 1-3 help to isolate the effect of chemical timescales observed in the steady-state detonation wave structure from the operator splitting perspective (although the degree to which the timescales are decreased is much higher in the simplified kinetics cases, see Fig.~\ref{fig:znd_comparison}). In these resolved regimes, the trends across cases 1-3 are consistent in both simplified and detailed mechanisms in that peak pressures and wave speeds observe small increases when dropping characteristic chemical timescales. However, these variations are small, and Fig.~\ref{fig:w_p_convergence} usefully reveals how chemical timescale modifications do not significantly alter convergence behavior in terms of large-scale detonation quantities. The fact that nearly identical trends in these macroscopic QoIs are observed in both simplified and detailed mechanisms points to the universality of this phenomenon.

\subsection{Convergence of Wave Structure}
\label{sec:wave_structure}

It is also important to investigate how chemical timescale reduction is reflected in alternations to the fine-scale (microscopic) detonation wave structure. As such, the goal of this section is to compare (a) instantaneous detonation wave profiles produced by the unsteady simulations at various resolutions, and (b) the degree to which the same modifications affect reaction zone locations in the detonation.

To this end, Fig.~\ref{fig:pressure_profiles} provides instantaneous pressure profiles at all tested grid resolutions for simplified and detailed kinetics configurations with corresponding ZND solutions overlaid. The profiles are plotted in the wave reference frame, with distance from the shock front $x_{s}$ normalized by the ZND induction length $L_{\text{IND}}$. As observed throughout Sec.~\ref{sec:macro}, initial inspection of Fig.~\ref{fig:pressure_profiles} reveals qualitatively similar trends across all cases, and provides more context to the macroscopic convergence trends observed in Fig.~\ref{fig:w_p_convergence}. Most notably, the fine-scale structures reveal how resolutions that achieve stabilization of detonation wave speeds at the ZND von-Neumann pressure do not necessarily imply complete recovery of an induction zone in unsteady simulations. More specifically, regardless of the kinetic mechanism used, the pressure profiles in Fig.~\ref{fig:pressure_profiles} reveal three characteristic regimes in detonation convergence.

\begin{figure}
    \centering
    \includegraphics[width=0.55\columnwidth]{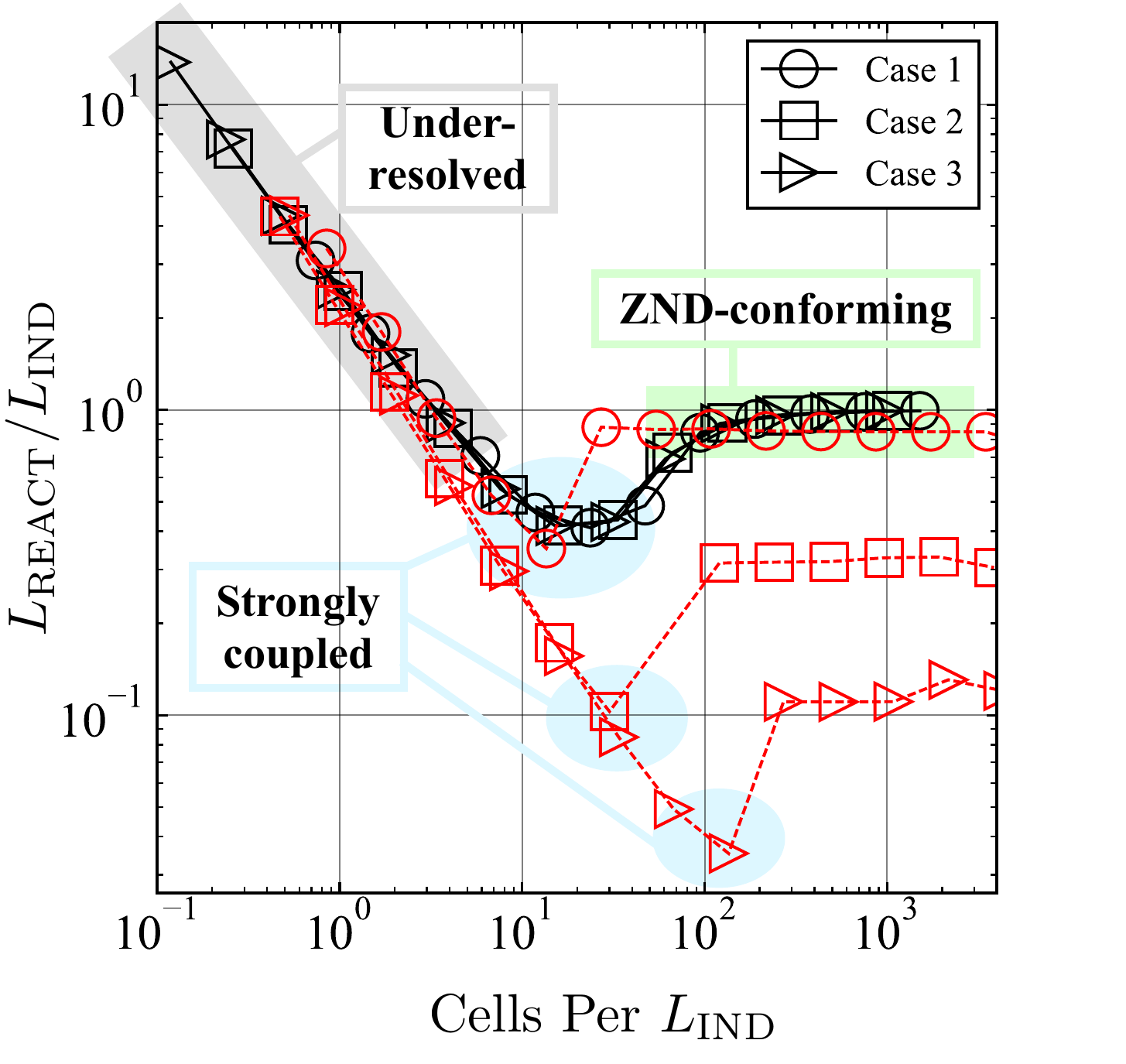}
    \caption{Convergence of detonation wave structure in terms of reaction front location for simplified (red) and detailed (black) kinetics configurations, with shaded regions indicating detonation convergence regimes. Y-axis is reaction front location normalized by ZND induction length, and X-axis is cells-per-ZND induction length (same as Fig.~\ref{fig:w_p_convergence}). Values are averages extracted from a set of detonation snapshots during steady propagation.}
    \label{fig:l_react_convergence}
\end{figure}

The first is the \textbf{under-resolved regime}, which is characterized by high variations in wave speed, reduced peak pressures, and a highly diffused detonation wave structure. The under-resolved regime is evident in Fig.~\ref{fig:pressure_profiles} through the simultaneous build-up of the pressure profiles towards the VN pressure and emergence of a non-diffuse shock wave, and ends in all cases at roughly 10 cells-per-induction length. 

The second is the \textbf{strongly-coupled regime}, which is characterized by stabilized wave speeds and roughly coincides with the point of peak pressure convergence, as shown in Fig.~\ref{fig:w_p_convergence}. Here, the CJ speed and VN pressures are reached and self-sustained, but there is very little distinction between the locations of peak chemical heat release and peak pressure. As such, the wave structure is still diffused through a smeared shock wave, and resembles the structures found in more complex 2D / 3D turbulence-influenced configurations. In Case 1 for the simplified kinetics model (Fig.~\ref{fig:pressure_profiles}(a)) and in all detailed kinetics cases (Fig.~\ref{fig:pressure_profiles}(c)), the strongly-coupled regime occurs near 10-50 cells-per-induction length. 

The third regime is the \textbf{ZND-conforming regime}, in which there is a clear formation of an induction zone and a pronounced separation between the reaction front and the leading shock wave. Transition from the strongly-coupled to ZND-conforming regime is visualized in Fig.~\ref{fig:pressure_profiles}(d), which showcases the emergence of an induction zone for Case 1 in the simplified kinetics configuration in the range of 20-to-50 cells-per-induction length (found to be slightly before the point at which the three detailed cases enter this regime, at roughly 100 cells-per-induction length). 

The effect of significant chemical timescale reduction (four orders of magnitude) on the fine-scale detonation wave structure is made clear when comparing Fig.~\ref{fig:pressure_profiles}(a) and (b), which respectively refer to cases 1 and 3 for the simplified kinetics configuration. Using this comparison, the chemical timescale reduction effect is present from two perspectives. First, in Fig.~\ref{fig:pressure_profiles}(b), the simulations appear to be exiting the strongly-coupled regime at a noticeably higher number of cells-per-induction length, meaning that more cells are required to converge in the highly-stiff case. Second, as the simulations exit the strongly-coupled regime, the developed induction zone in Fig.~\ref{fig:pressure_profiles}(b) converges to a length much smaller than what is observed in the corresponding steady-state ZND profile, resulting in qualitative departure from Case 1 and the three detailed cases. In other words, the drastic reduction in chemical timescale in the simplified configurations is preventing a complete separation between reaction and shock fronts, and necessitates a closer inspection into the convergence of reaction front locations observed in the unsteady simulations.

As such, Fig.~\ref{fig:l_react_convergence} shows the distance between the wave front and the primary reaction zone ($L_{\text{REACT}}$), providing a type of global diagram for the three regimes described above. Here, the wave front location is identified by the first cell to exceed the ambient pressure, while the reaction front location is identified by the cell with the highest volumetric heat release rate. Note that in the classical ZND model, this distance is referred to as the induction length; however, this distinction is made in the context of Fig.~\ref{fig:l_react_convergence} because the coarser simulations lack a typical induction zone. As was seen in the instantaneous profiles in Fig.~\ref{fig:pressure_profiles}, the reaction zone sits closer to the shock front as the grid is refined in the under-resolved and strongly-coupled regimes. However, once the ZND-conforming regime is reached, the reaction front shifts back from the shock front as a result of the formation of a clear induction zone.

Figure~\ref{fig:l_react_convergence} illuminates several interesting qualities about detonation convergence from the reaction front perspective. First, normalization by induction length collapses both simplified (Case 1) and all detailed mechanism curves, particularly in the under-resolved region. Second, movement out of under-resolved and into strongly-coupled detonations occurs at a nearly fixed convergence rate with respect to detonation wave structure resolution. Additionally, the transition out of the strongly-coupled regime into the ZND-conforming regime occurs less smoothly for the simplified Case 1 setting, as opposed to all three detailed cases. Perhaps most apparent in Fig.~\ref{fig:l_react_convergence} is that there is negligible difference in the curves corresponding to the detailed cases, whereas cases 2 and 3 in the simplified configurations reflect the trends observed in the comparison between Fig.~\ref{fig:pressure_profiles}(a) and (b). Namely, the simplified cases 2 and 3, which successively drop the chemical timescale relative to case 1, do not converge to the steady-state baseline wave structure. 

This finding illuminates how deviation from the steady-state solution in the ZND-conforming regime—which denotes a significant departure from ``standard" detonation convergence properties—increases not solely with a reduction in chemical timescales, but with a reduction in the ratio of {chemical timescales to induction times}, {where the induction time is defined as the integrated inverse local fluid velocity in the induction zone (see caption in Table~\ref{table:cases})}. This ratio is provided in the last row of Table~\ref{table:cases}. Note that through activation energy modifications, the simplified cases 1 through 3 reduce both chemical timescale and this ratio. Meanwhile, ambient pressure increases in the the detailed cases reduce the chemical timescales, but not the ratio of {chemical timescale to induction time}.

It is hypothesized that the deviation in the ZND-conforming regime observed by cases 2 and 3 in the simplified kinetics configuration is a direct consequence of (a) operator splitting errors due to the global time integration formulation, (b) chemical time integration errors that arise in highly stiff regimes, or (c) a combination of the two. In any case, since the timescale effects are not as pronounced in the detailed chemistry case, the trends in Fig.~\ref{fig:l_react_convergence} highlight the importance of the ratio of the chemical timescale to the {induction times} in assessment of detonation convergence. The practical implication here is that, if the {induction time} can be maintained (i.e., if induction lengths and CJ speeds are not significantly altered), a reduction in chemical timescales due to mechanism simplifications and re-designs via combustion modeling can result in vastly improved detonation convergence properties. 

\section{Conclusion}
\label{sec:conclusion}
The goal of this work was to isolate effects of chemical timescale reductions on the convergence of one-dimensional channel detonation simulations in the context of operator splitting. An AMR-based solver was used to this end, which enabled resolutions up to $\mathcal{O}(1000)$ cells-per-induction length in unsteady detonations. The analysis was conducted using two chemical kinetics configurations: (1) a simplified three-step model mechanism, in which chemical timescales in the detonation were modified by adjusting activation energies, and (2) a detailed hydrogen mechanism utilizing a 2H$_2$:O$_2$:7Ar reactant composition, in which chemical timescales were adjusted through ambient pressure modifications. 

Chemical timescale reductions had negligible effect on large-scale macroscopic (peak pressure and wave speed) convergence characteristics for the cases considered here. Convergence of small-scale quantities was then assessed using instantaneous detonation profiles and relative locations of reaction zones in the wave structure. Supplementing the macroscopic convergence trends, the wave structure analysis revealed three characteristic detonation convergence regimes, namely the under-resolved, strongly-coupled, and ZND-conforming regimes. Convergence of reaction front locations uncovered the crucial role played by not only the chemical timescale, but also the ratio of {chemical timescale to induction time}. The simplified cases exhibiting drastic reductions to this ratio showed not only delayed convergence to the strongly-coupled regime, but also significantly more deviation from baseline ZND solutions. The practical implication here is that if the {induction time} can be maintained, a reduction in chemical timescales due to mechanism simplifications and re-designs via combustion modeling can result in vastly improved detonation convergence properties. {Ultimately, this work highlights not only the important role played by the chemical timescale in detonation convergence, but also reveals the utility in assessing detonation convergence from the perspective of reaction front locations.}

{The trends uncovered here open several avenues for extended verification and future work. A natural extension to the current study is to perform the same analysis, but for different fuels and mechanism settings. For example, investigating similar trends in convergence properties under chaotic and pulsating detonation simulation regimes (such as those uncovered in Ref.~\cite{ng2005nonlinear}) is a promising direction. Additionally, a direct way to build on the current study is to explore how higher spatial dimensions modify the convergence regimes, and the extent to which chemical timescale modifications can affect the cellular detonation structure. Considering that different cellular patterns may be obtained for different initial perturbations of the detonation wave, convergence of cell sizes may only be obtained in a statistical sense from an ensemble of simulations at the same resolution \cite{sharpe2011_cellstat}. Even with the computational acceleration enabled by AMR, such an analysis would likely be extremely computationally expensive, especially when done in conjunction with investigations of chemical stiffness effects. In lieu of this, repeating the one-dimensional analysis in this study with coupled time integration schemes—instead of operator-split schemes—that attempt to remove splitting errors would lead to useful insights concerning the role played by chemical stiffness on the nature of detonation convergence. Such an analysis can build on previous work that has explored the relationship splitting errors and numerical solution deviations on different combustion problems \cite{lu2017analysis}.}

Further, the framework in this study provides a methodology for mechanism reduction. Since shock-capturing requires somewhat stringent requirements as compared to subsonic flames, mechanism reduction could now focus on stiffness reduction within this constraint. In fact, related work suggests that machine learning models can be trained to enable precisely such approaches \cite{vijay}. 

\section*{Declaration of competing interest} 
The authors declare that they have no known competing financial interests or personal relationships that could have appeared to influence the work reported in this paper.

\section*{Acknowledgments} 
The authors acknowledge funding through ONR MURI N00014-22-1-2606, with Dr. Steven Martens as program manager. SB acknowledges support from Argonne National Laboratory under DOE contract DE-AC02-06CH11357.

\bibliography{references}

\begin{thebibliography}{10}
\expandafter\ifx\csname url\endcsname\relax
  \def\url#1{\texttt{#1}}\fi
\expandafter\ifx\csname urlprefix\endcsname\relax\def\urlprefix{URL }\fi
\expandafter\ifx\csname href\endcsname\relax
  \def\href#1#2{#2} \def\path#1{#1}\fi

\bibitem{sharma2024amrex}
S.~Sharma, R.~Bielawski, O.~Gibson, S.~Zhang, V.~Sharma, A.~H. Rauch, J.~Singh, S.~Abisleiman, M.~Ullman, S.~Barwey, et~al., An amrex-based compressible reacting flow solver for high-speed reacting flows relevant to hypersonic propulsion, arXiv preprint arXiv:2412.00900 (2024).

\bibitem{short1997nonlinear}
M.~Short, J.~J. Quirk, On the nonlinear stability and detonability limit of a detonation wave for a model three-step chain-branching reaction, Journal of Fluid Mechanics 339 (1997) 89--119.

\bibitem{shepherd_h2air_web}
J.~Shepherd, {Chemical Reaction and Thermodynamic Data - Cantera Format}, explosion Dynamics Laboratory, California Institute of Technology (2018).

\bibitem{poludnenko2019unified}
A.~Y. Poludnenko, J.~Chambers, K.~Ahmed, V.~N. Gamezo, B.~D. Taylor, A unified mechanism for unconfined deflagration-to-detonation transition in terrestrial chemical systems and type ia supernovae, Science 366~(6465) (2019) eaau7365.

\bibitem{wang_2023}
J.~Crane, J.~T. Lipkowicz, X.~Shi, I.~Wlokas, A.~M. Kempf, H.~Wang, Three-dimensional detonation structure and its response to confinement, Proceedings of the Combustion Institute 39~(3) (2023) 2915--2923.

\bibitem{meagher2023isolating}
P.~A. Meagher, X.~Shi, J.~P. Santos, N.~K. Muraleedharan, J.~Crane, A.~Y. Poludnenko, H.~Wang, X.~Zhao, Isolating gasdynamic and chemical effects on the detonation cellular structure: A combined experimental and computational study, Proceedings of the Combustion Institute 39~(3) (2023) 2865--2873.

\bibitem{venkat_rde_arfm}
V.~Raman, S.~Prakash, M.~Gamba, Nonidealities in rotating detonation engines, Annual Review of Fluid Mechanics 55 (2023) 639--674.

\bibitem{supraj_liquid}
S.~Prakash, R.~Bielawski, V.~Raman, K.~Ahmed, J.~Bennewitz, Three-dimensional numerical simulations of a liquid rp-2/o2 based rotating detonation engine, Combustion and Flame 259 (2024) 113097.

\bibitem{gutmark2019_pecs}
V.~Anand, E.~Gutmark, Rotating detonation combustors and their similarities to rocket instabilities, Progress in Energy and Combustion Science 73 (2019) 182--234.

\bibitem{wolanski2013_pci}
P.~Wola{\'n}ski, Detonative propulsion, Proceedings of the Combustion Institute 34~(1) (2013) 125--158.

\bibitem{short_arfm_2018}
M.~Short, J.~J. Quirk, High explosive detonation--confiner interactions, Annual Review of Fluid Mechanics 50 (2018) 215--242.

\bibitem{voelkel2022effect}
S.~J. Voelkel, E.~K. Anderson, M.~Short, C.~Chiquete, S.~I. Jackson, Effect of lot microstructure variations on detonation performance of the triaminotrinitrobenzene (tatb)-based insensitive high explosive pbx 9502, Combustion and Flame 246 (2022) 112373.

\bibitem{yanez2015analysis}
J.~Yanez, M.~Kuznetsov, A.~Souto-Iglesias, An analysis of the hydrogen explosion in the fukushima-daiichi accident, International Journal of Hydrogen Energy 40~(25) (2015) 8261--8280.

\bibitem{ng2007hazard}
H.~D. Ng, Y.~Ju, J.~H. Lee, Assessment of detonation hazards in high-pressure hydrogen storage from chemical sensitivity analysis, International Journal of Hydrogen Energy 32~(1) (2007) 93--99.

\bibitem{powers_paolucci}
J.~M. Powers, S.~Paolucci, Accurate spatial resolution estimates for reactive supersonic flow with detailed chemistry, AIAA journal 43~(5) (2005) 1088--1099.

\bibitem{mazaheri2007effect}
K.~MAZAHERI, S.~Hashemi, The effect of chain initiation reaction on the stability of gaseous detonations, Combustion Science and Technology 179~(8) (2007) 1701--1736.

\bibitem{ng2005numerical}
H.~Ng, M.~Radulescu, A.~Higgins, N.~Nikiforakis, J.~Lee, Numerical investigation of the instability for one-dimensional chapman--jouguet detonations with chain-branching kinetics, Combustion Theory and Modelling 9~(3) (2005) 385--401.

\bibitem{yungster_ctm_2004}
S.~Yungster, K.~Radhakrishnan, Pulsating one-dimensional detonations in hydrogen--air mixtures, Combustion Theory and Modelling 8~(4) (2004) 745.

\bibitem{han2019pulsation}
W.~Han, C.~Wang, C.~K. Law, Pulsation in one-dimensional h2--o2 detonation with detailed reaction mechanism, Combustion and Flame 200 (2019) 242--261.

\bibitem{schwer_kailasanath}
D.~Schwer, K.~Kailasanath, Numerical investigation of the physics of rotating-detonation-engines, Proceedings of the Combustion Institute 33~(2) (2011) 2195--2202.

\bibitem{mazaheri2012diffusion}
K.~Mazaheri, Y.~Mahmoudi, M.~I. Radulescu, Diffusion and hydrodynamic instabilities in gaseous detonations, Combustion and Flame 159~(6) (2012) 2138--2154.

\bibitem{ebrahimi_pulse}
H.~B. Ebrahimi, C.~L. Merkle, Numerical simulation of a pulse detonation engine with hydrogen fuels, Journal of Propulsion and Power 18~(5) (2002) 1042--1048.

\bibitem{musick2023numerical}
B.~J. Musick, M.~Paudel, P.~K. Ramaprabhu, J.~A. McFarland, Numerical simulations of droplet evaporation and breakup effects on heterogeneous detonations, Combustion and Flame 257 (2023) 113035.

\bibitem{meng2021distributions}
Q.~Meng, N.~Zhao, H.~Zhang, On the distributions of fuel droplets and in situ vapor in rotating detonation combustion with prevaporized n-heptane sprays, Physics of Fluids 33~(4) (2021).

\bibitem{michael_stratification}
M.~Ullman, S.~Prakash, S.~Barwey, V.~Raman, Detonation structure in the presence of mixture stratification using reaction-resolved simulations, Combustion and Flame 264 (2024) 113427.

\bibitem{ryu2024numerical}
J.~I. Ryu, X.~Shi, J.-Y. Chen, Numerical investigation of detonation propagation through fuel-stratified layers, Proceedings of the Combustion Institute 40~(1-4) (2024) 105510.

\bibitem{suzuki2024dns}
S.~Suzuki, K.~Iwata, R.~Kai, R.~Kurose, A dns study of detonation in h2/o2 mixture with variable-intensity turbulences, Proceedings of the Combustion Institute 40~(1-4) (2024) 105337.

\bibitem{machida2014ddt}
T.~Machida, M.~Asahara, A.~K. Hayashi, N.~Tsuboi, Three-dimensional simulation of deflagration-to-detonation transition with a detailed chemical reaction model, Combustion Science and Technology 186~(10-11) (2014) 1758--1773.

\bibitem{ramachandran2023ddt}
S.~Ramachandran, N.~Srinivasan, Z.~Wang, A.~Behkish, S.~Yang, A numerical investigation of deflagration propagation and transition to detonation in a microchannel with detailed chemistry: Effects of thermal boundary conditions and vitiation, Physics of Fluids 35~(7) (2023).

\bibitem{han2019three}
W.~Han, C.~Wang, C.~K. Law, Three-dimensional simulation of oblique detonation waves attached to cone, Physical Review Fluids 4~(5) (2019) 053201.

\bibitem{abisleiman2024high}
S.~Abisleiman, R.~Bielawski, V.~Raman, High-fidelity simulation of oblique detonation waves, in: AIAA SCITECH 2024 Forum, 2024, p. 1656.

\bibitem{debnath2023computational}
P.~Debnath, K.~M. Pandey, Computational analysis of shrouded ejector effect on starting vortex and combustion efficiency in pulse detonation combustor with different fuels, Combustion Science and Technology (2023) 1--23.

\bibitem{debnath2023numerical}
P.~Debnath, K.~M. Pandey, Numerical analysis on detonation wave and combustion efficiency of pulse detonation combustor with u-shape combustor, Journal of Thermal Science and Engineering Applications 15~(10) (2023) 101006.

\bibitem{takuma_proci}
T.~Sato, F.~Chacon, L.~White, V.~Raman, M.~Gamba, Mixing and detonation structure in a rotating detonation engine with an axial air inlet, Proceedings of the Combustion Institute 38~(3) (2021) 3769--3776.

\bibitem{pinaki_les_rde}
P.~Pal, J.~Braun, Y.~Wang, V.~Athmanathan, G.~Paniagua, T.~R. Meyer, Large-eddy simulation study of flow and combustion dynamics in a full-scale hydrogen--air rotating detonation combustor-stator integrated system, Journal of Engineering for Gas Turbines and Power 147~(3) (2025) 031002.

\bibitem{ullman2023_drone}
M.~Ullman, S.~Prakash, D.~Jackson, V.~Raman, C.~Slabaugh, J.~Bennewitz, Self-excited wave stabilization in a linear detonation combustor, Combustion and Flame 257 (2023) 113044.

\bibitem{menon_2024}
M.~Salvadori, A.~Panchal, S.~Menon, Simulation of wave mode switching in a rotating detonation engine with gaseous and liquid fuel, Aerospace Science and Technology 147 (2024) 109008.

\bibitem{supraj_linearized}
S.~Prakash, R.~Fi{\'e}vet, V.~Raman, J.~Burr, K.~H. Yu, Analysis of the detonation wave structure in a linearized rotating detonation engine, AIAA Journal 58~(12) (2020) 5063--5077.

\bibitem{supraj_rde_proci}
S.~Prakash, V.~Raman, C.~F. Lietz, W.~A. Hargus~Jr, S.~A. Schumaker, Numerical simulation of a methane-oxygen rotating detonation rocket engine, Proceedings of the Combustion Institute 38~(3) (2021) 3777--3786.

\bibitem{rosato2021stabilized}
D.~A. Rosato, M.~Thornton, J.~Sosa, C.~Bachman, G.~B. Goodwin, K.~A. Ahmed, Stabilized detonation for hypersonic propulsion, Proceedings of the national academy of sciences 118~(20) (2021) e2102244118.

\bibitem{detfoam_paper}
R.~Bielawski, S.~Barwey, S.~Prakash, V.~Raman, Highly-scalable gpu-accelerated compressible reacting flow solver for modeling high-speed flows, Computers \& Fluids (2023) 105972.

\bibitem{hu2017numerical}
G.~Hu, A numerical study of 2d detonation waves with adaptive finite volume methods on unstructured grids, Journal of Computational Physics 331 (2017) 297--311.

\bibitem{bourlioux_ppm_1991}
A.~Bourlioux, A.~J. Majda, V.~Roytburd, Theoretical and numerical structure for unstable one-dimensional detonations, SIAM Journal on Applied Mathematics 51~(2) (1991) 303--343.

\bibitem{dg_detonation}
J.~Du, C.~Wang, C.~Qian, Y.~Yang, High-order bound-preserving discontinuous galerkin methods for stiff multispecies detonation, SIAM Journal on Scientific Computing 41~(2) (2019) B250--B273.

\bibitem{fedkiw1999ghost}
R.~P. Fedkiw, T.~Aslam, S.~Xu, The ghost fluid method for deflagration and detonation discontinuities, Journal of Computational Physics 154~(2) (1999) 393--427.

\bibitem{blanquart_jcp_2015}
B.~Savard, Y.~Xuan, B.~Bobbitt, G.~Blanquart, A computationally-efficient, semi-implicit, iterative method for the time-integration of reacting flows with stiff chemistry, Journal of Computational Physics 295 (2015) 740--769.

\bibitem{qian2020convergence}
C.~Qian, C.~Wang, J.~Liu, A.~Brandenburg, N.~E. Haugen, M.~A. Liberman, Convergence properties of detonation simulations, Geophysical \& Astrophysical Fluid Dynamics 114~(1-2) (2020) 58--76.

\bibitem{mazaheri_thesis}
B.~K. Mazaheri, Mechanism of the onset of detonation in blast initiation, Ph.D. thesis, McGill University (2000).

\bibitem{yungster2005structure}
S.~Yungster, K.~Radhakrishnan, Structure and stability of one-dimensional detonationsin ethylene-air mixtures, Shock Waves 14~(1-2) (2005) 61--72.

\bibitem{zhao2021pulsating}
M.~Zhao, Z.~Ren, H.~Zhang, Pulsating detonative combustion in n-heptane/air mixtures under off-stoichiometric conditions, Combustion and Flame 226 (2021) 285--301.

\bibitem{strang}
G.~Strang, On the construction and comparison of difference schemes, SIAM Journal on Numerical Analysis 5~(3) (1968) 506--517.

\bibitem{hllc}
P.~Batten, N.~Clarke, C.~Lambert, D.~M. Causon, On the choice of wavespeeds for the hllc riemann solver, SIAM Journal on Scientific Computing 18~(6) (1997) 1553--1570.

\bibitem{AMReX_JOSS}
W.~Zhang, A.~Almgren, V.~Beckner, J.~Bell, J.~Blaschke, C.~Chan, M.~Day, B.~Friesen, K.~Gott, D.~Graves, M.~Katz, A.~Myers, T.~Nguyen, A.~Nonaka, M.~Rosso, S.~Williams, M.~Zingale, {AMReX}: a framework for block-structured adaptive mesh refinement, Journal of Open Source Software 4~(37) (2019) 1370.

\bibitem{ng2003direct}
H.~D. Ng, J.~H. Lee, Direct initiation of detonation with a multi-step reaction scheme, Journal of Fluid Mechanics 476 (2003) 179--211.

\bibitem{shepherd_h2air_3}
R.~M{\'e}vel, S.~Javoy, F.~Lafosse, N.~Chaumeix, G.~Dupr{\'e}, C.-E. Paillard, Hydrogen--nitrous oxide delay times: Shock tube experimental study and kinetic modelling, Proceedings of the Combustion Institute 32~(1) (2009) 359--366.

\bibitem{radulescu2002_argon}
M.~I. Radulescu, H.~D. Ng, J.~H. Lee, B.~Varatharajan, The effect of argon dilution on the stability of acetylene/oxygen detonations, Proceedings of the Combustion Institute 29~(2) (2002) 2825--2831.

\bibitem{sdtoolbox}
S.~Browne, J.~Ziegler, N.~Bitter, B.~Schmidt, J.~Lawson, J.~Shepherd, Sdtoolbox: Numerical tools for shock and detonation wave modeling, Explosion Dynamics Laboratory. GALCIT Report FM2018 1 (2018).

\bibitem{ng2005nonlinear}
H.~Ng, A.~Higgins, C.~Kiyanda, M.~Radulescu, J.~Lee, K.~Bates, N.~Nikiforakis, Nonlinear dynamics and chaos analysis of one-dimensional pulsating detonations, Combustion Theory and Modelling 9~(1) (2005) 159--170.

\bibitem{sharpe2011_cellstat}
G.~Sharpe, M.~Radulescu, Statistical analysis of cellular detonation dynamics from numerical simulations: One-step chemistry, Combustion Theory and Modelling 15~(5) (2011) 691--723.

\bibitem{lu2017analysis}
Z.~Lu, H.~Zhou, S.~Li, Z.~Ren, T.~Lu, C.~K. Law, Analysis of operator splitting errors for near-limit flame simulations, Journal of Computational Physics 335 (2017) 578--591.

\bibitem{vijay}
V.~Vijayarangan, H.~A. Uranakara, S.~Barwey, R.~M. Galassi, M.~R. Malik, M.~Valorani, V.~Raman, H.~G. Im, A data-driven reduced-order model for stiff chemical kinetics using dynamics-informed training, Energy and AI (2023) 100325.

\end{thebibliography}

\end{document}